\documentclass{appolb}
\usepackage{epsfig}

\begin{document}
\title{Effect of temperature-dependent $\eta/s$ on flow
       anisotropies\thanks{Parallel talk at SQM2011, Krakow}%
}
\author{H.\ Niemi${}^{a,b}$, G.S.\ Denicol${}^{c}$, P.\ Huovinen${}^{c}$, 
E.\ Moln\'ar${}^{d}$, and\\ D.H.\ Rischke${}^{a,c}$
\address{$^{a}$Frankfurt Institute for Advanced Studies,
\\ Ruth-Moufang-Str.\ 1, D-60438 Frankfurt am Main, Germany\\
$^{b}$Department of Physics, P.O.Box 35, FI-40014 University of Jyv\"askyl\"a,
Finland\\
$^{c}$Institut f\"ur Theoretische Physik, Johann Wolfgang
Goethe-Universit\"at, Max-von-Laue-Str.\ 1, D-60438 Frankfurt am Main, Germany\\
$^{d}$MTA-KFKI, Research Institute for Particle and Nuclear Physics,
\\ H-1525 Budapest, P.O.Box 49, Hungary}
}
\maketitle
\begin{abstract}
We investigate the effects of a temperature-dependent shear viscosity
over entropy density ratio $\eta/s$ on the flow anisotropy
coefficients $v_2$ and $v_4$ in ultrarelativistic heavy-ion collisions
at RHIC and LHC. We find that $v_4$ is more sensitive to the viscosity
at low temperatures than $v_2$. At RHIC $v_2$ is mostly affected by
the viscosity around the phase transition, but the larger the
collision energy, the more the quark-gluon plasma viscosity
affects $v_2$.
\end{abstract}
  
\section{Introduction}

Presently, most works aiming at the determination of the shear
viscosity of strongly interacting matter assume a constant shear
viscosity over entropy density ratio, $\eta/s$. However, this ratio
can be a strongly varying function of temperature both in hadronic
matter and in the quark-gluon plasma. In this work we study
consequences of such a temperature dependence~\cite{Niemi:2011ix}.

We model the space-time evolution of matter formed in heavy-ion
collisions using relativistic dissipative hydrodynamics~\cite{IS}.
We assume longitudinal boost invariance and neglect the net-baryon
number. Essential inputs to the model are the equation of state, the
initial state and the transport coefficients.  We consider here only
the shear viscosity.

As equation of state we use a recent lattice
parametrisation~\cite{Huovinen:2009yb} with chemical freeze-out at
$T = 150$ MeV. The initial energy density at $\tau_0 = 1.0$~fm is
proportional to the density of binary nucleon-nucleon collisions in
the transverse plane. The maximum energy density is fixed to reproduce
the measured multiplicity in the most central
collisions~\cite{Adler:2003cb, Aamodt:2010pb}. For 
$\sqrt{s_{NN}}=5.5$ TeV Pb+Pb collisions we use the multiplicity
predicted by the minijet + saturation model~\cite{Eskola:2005ue}.
To compensate for different entropy production for different
parametrizations of the shear viscosity, the initial energy-density
profiles are normalised differently for each parametrization.
Freeze-out is implemented using the Cooper-Frye
formula~\cite{Cooper:1974mv} on a $T_{\rm dec} = 100$ MeV hypersurface
including the dissipative correction $\delta f$ to the thermal
distributions.

\begin{figure}[t]
 \begin{center}
  \includegraphics[width=6.0cm]{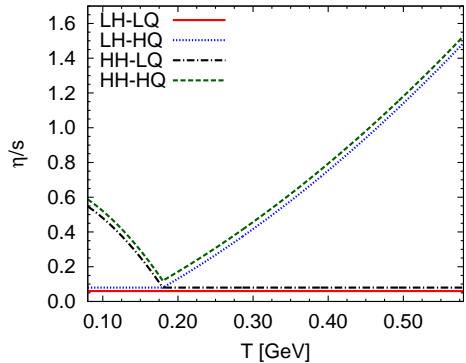}
  \caption{$\eta/s$ parametrizations.}
  \label{fig:etapers}
 \end{center}
\end{figure}
For $\eta/s$, we consider the four different parametrizations shown in
Fig.~\ref{fig:etapers}. The minimum value of $\eta/s$ is fixed to be
$\eta/s=0.08$ at $T=180$ MeV for all parametrizations. The shear
relaxation time~\cite{IS} is taken to be $\tau_R = 5\eta/(e+p)$, where
$e$ and $p$ are energy density and pressure, respectively.

\section{Results}

\begin{figure}[t]
\begin{minipage}[b]{60mm}
\includegraphics[width=58mm]{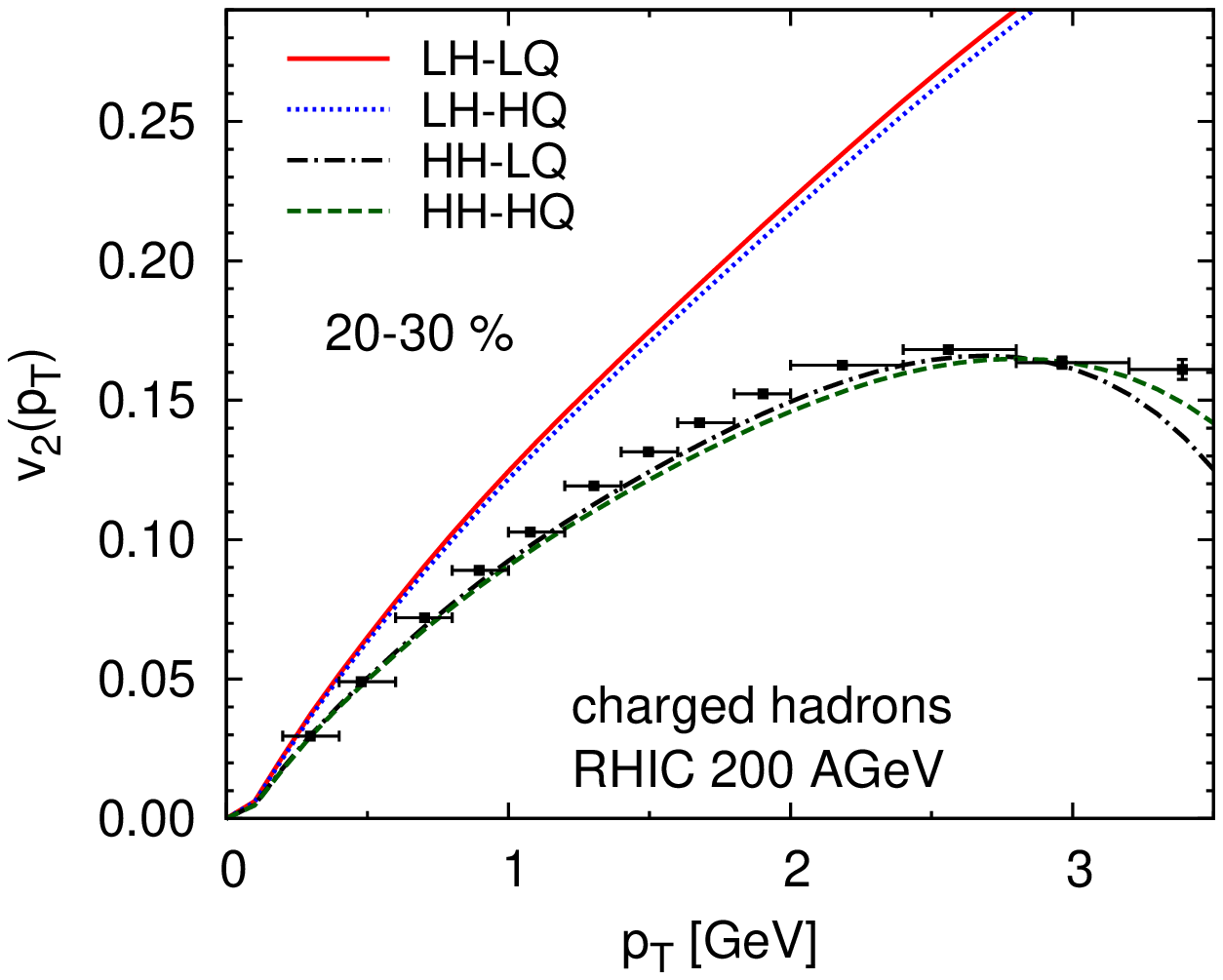}
\end{minipage}
 \hfill
\begin{minipage}[b]{63mm}
 \includegraphics[width=6.3cm]{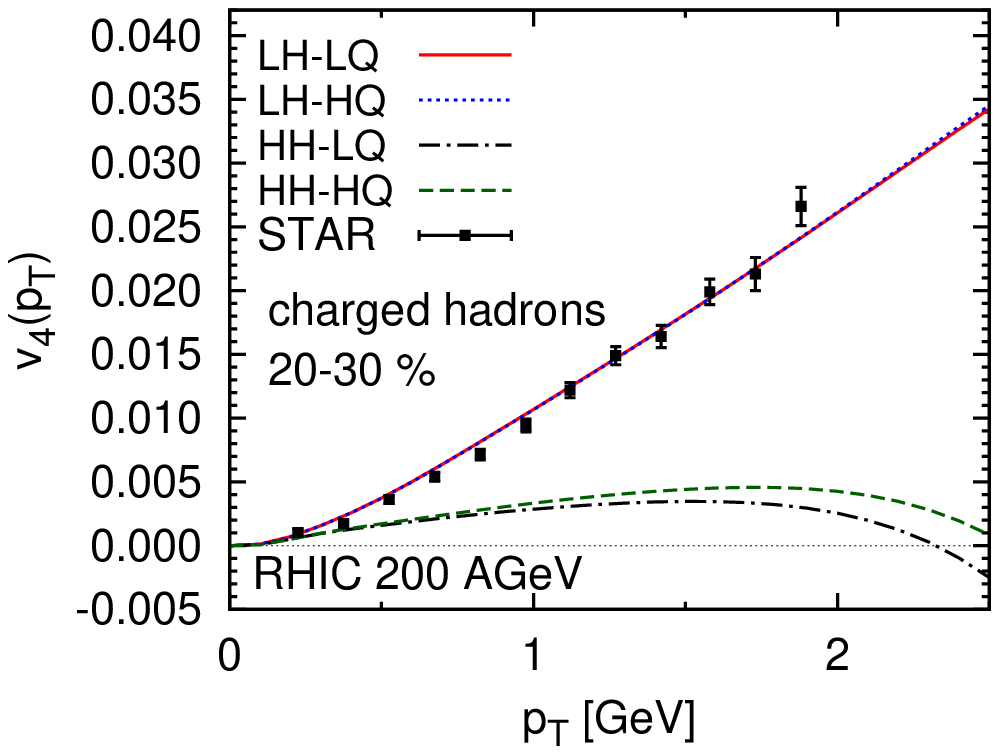}
\end{minipage}
\caption{$v_2(p_T)$ (left) and $v_4(p_T)$ (right) of charged hadrons in
the 20-30\% most central Au+Au collisions at $\sqrt{s_{NN}} = 200$ GeV (RHIC).
Data are from Refs.~\cite{Bai,v4}.}
 \label{fig:vnRHIC}
\end{figure}

\begin{figure}[b]
\begin{minipage}[b]{60mm}
 \includegraphics[width=58mm]{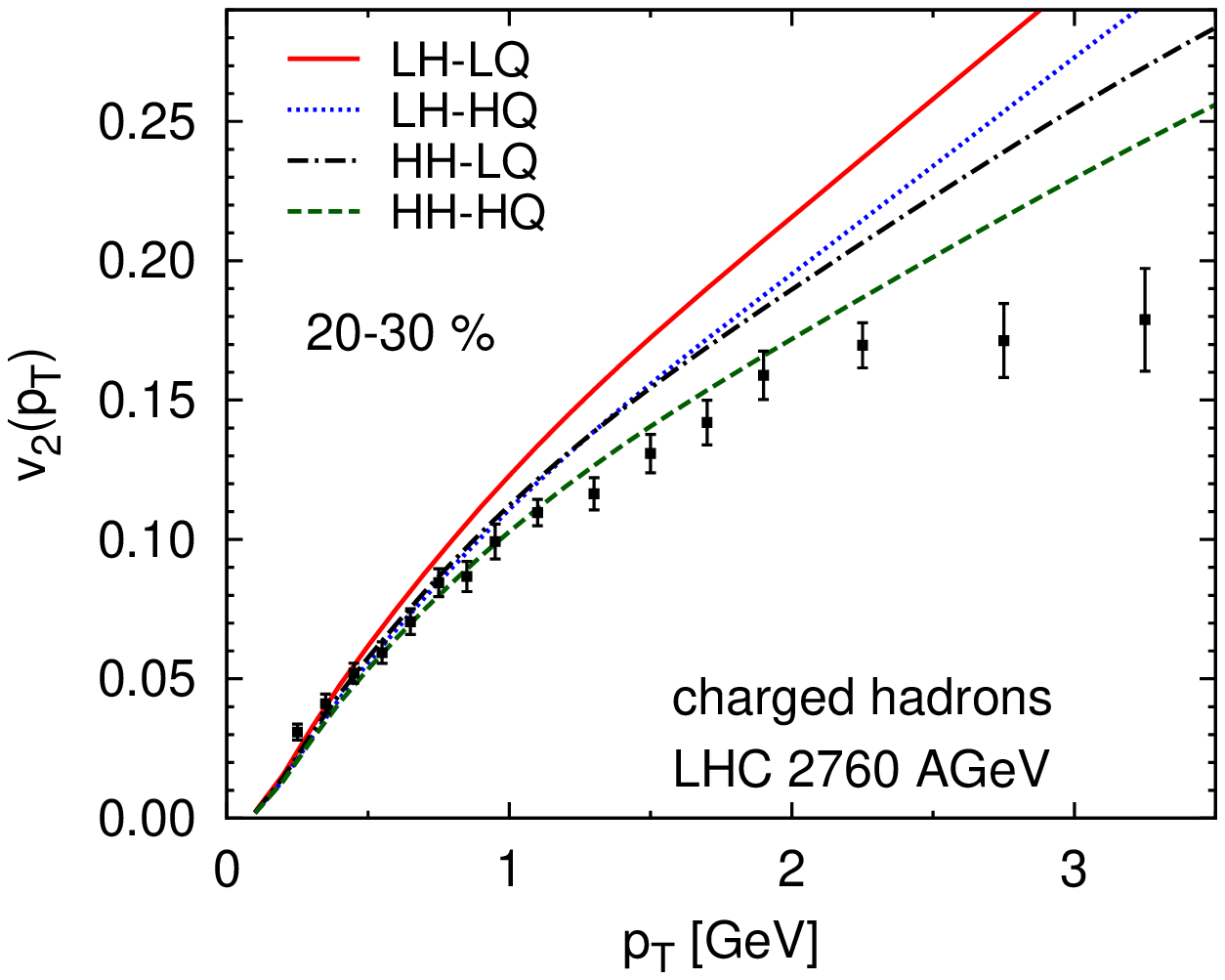}
\end{minipage}
 \hfill
\begin{minipage}[b]{63mm}
 \includegraphics[width=6.3cm]{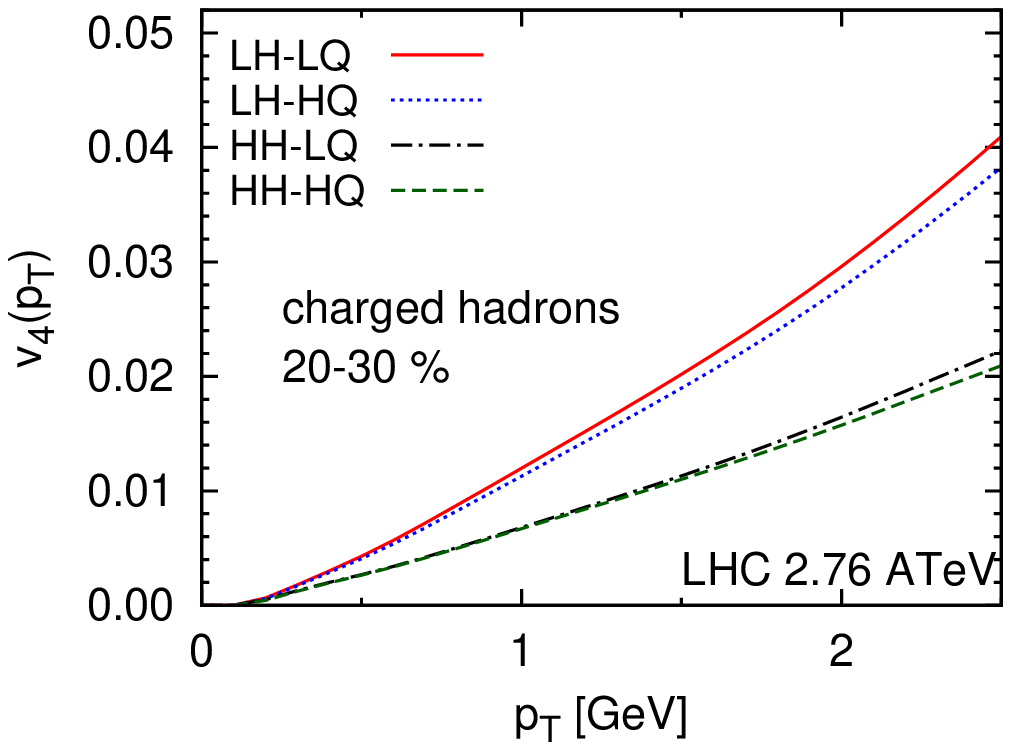}
\end{minipage}
\caption{$v_2(p_T)$ (left) and $v_4(p_T)$ (right) of charged hadrons in
the 20-30\% most central Pb+Pb collisions at $\sqrt{s_{NN}} = 2.76$ TeV (LHC).
Data are from Ref.~\cite{Aamodt:2010pa}.}
 \label{fig:vnLHClow}
\end{figure}

\begin{figure}[t]
\begin{minipage}[b]{60mm}
\includegraphics[width=58mm]{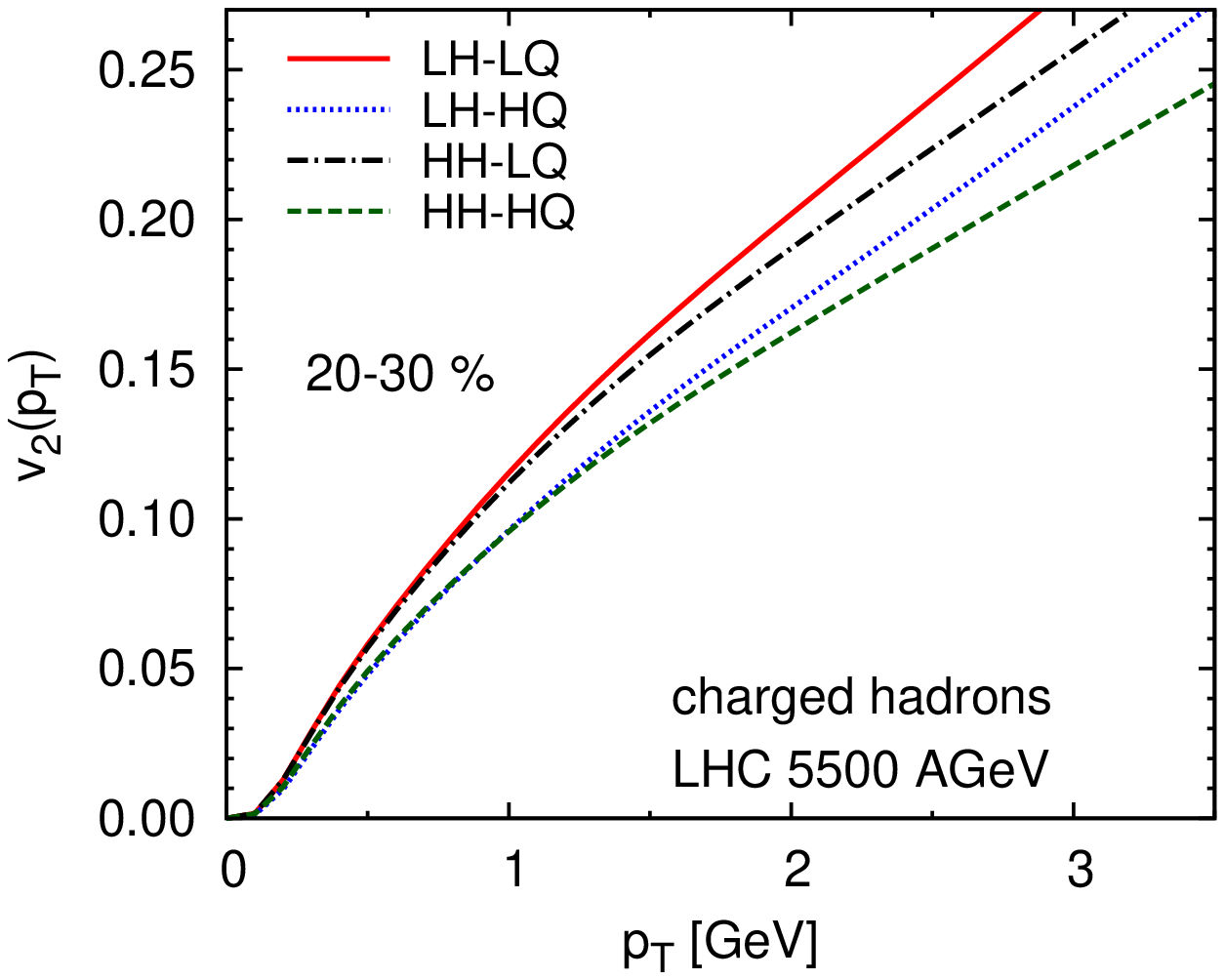}
\end{minipage}
 \hfill
\begin{minipage}[b]{63mm}
 \includegraphics[width=6.3cm]{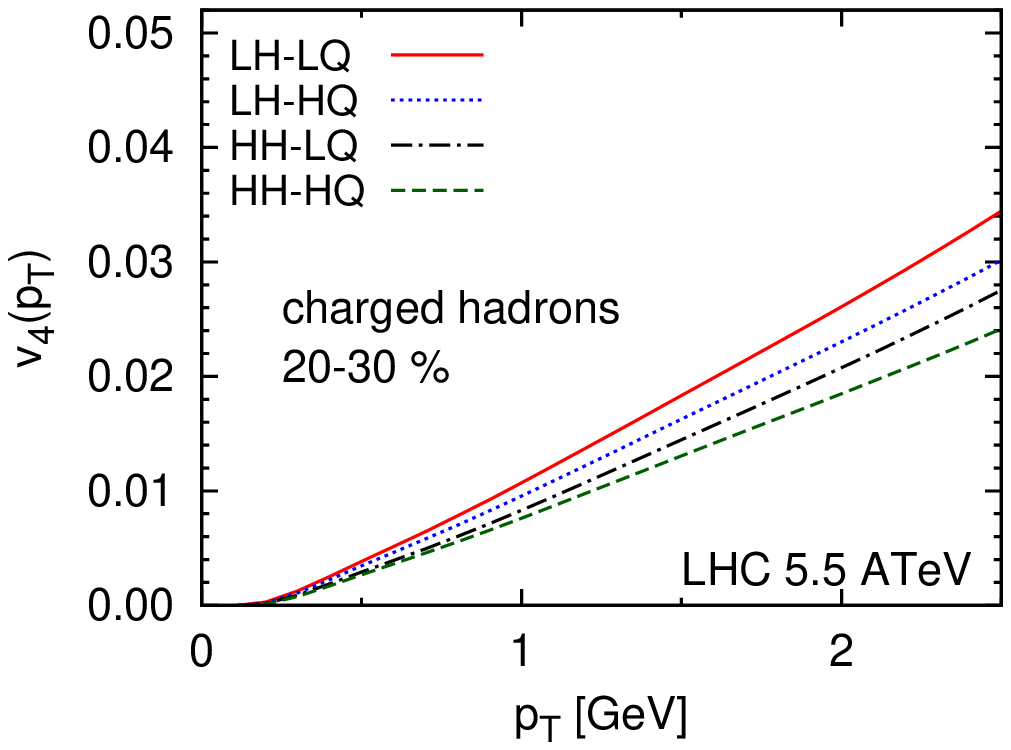}
\end{minipage}
\caption{$v_2(p_T)$ (left) and $v_4(p_T)$ (right) of charged hadrons in
the 20-30\% most central Pb+Pb collisions at $\sqrt{s_{NN}} = 5.5$ TeV (LHC).}
 \label{fig:vnLHCfull}
\end{figure}

The elliptic flow coefficient, $v_2(p_T)$, for charged hadrons in the
20-30\% most central collisions at RHIC, at the present LHC energy,
and at the full LHC collision energy is shown in the left panels of
Figs.~\ref{fig:vnRHIC}--\ref{fig:vnLHCfull}, respectively. We note
that at RHIC, the high-temperature part of $\eta/s$ has practically no
effect on the results. On the other hand, the viscous suppression of
$v_2(p_T)$ is strongly enhanced if we increase the hadronic
$\eta/s$. In low-energy LHC collisions, both hadronic and QGP
viscosity have a similar effect, whereas at the full LHC energy the
behaviour is opposite to that seen at RHIC: $v_2(p_T)$ is almost
independent of the hadronic $\eta/s$, but sensitive to the
high-temperature viscosity.

The anisotropy coefficient $v_4(p_T)$ (right panels of
Figs.~\ref{fig:vnRHIC}--\ref{fig:vnLHCfull}) exhibits a similar
pattern where the sensitivity to the low-temperature viscosity
decreases, and the sensitivity to the high-temperature viscosity
increases with increasing collision energy. However, in general
$v_4(p_T)$ is more sensitive to the shear viscosity in the
low-temperature region than $v_2(p_T)$. At the low LHC energy
$v_4(p_T)$ behaves like $v_2(p_T)$ at RHIC, and at the full LHC energy
$v_4(p_T)$ behaves like $v_2(p_T)$ at low LHC energy.

\begin{figure}[b]
 \begin{minipage}[t]{55mm}
  \includegraphics[width=5.1cm]{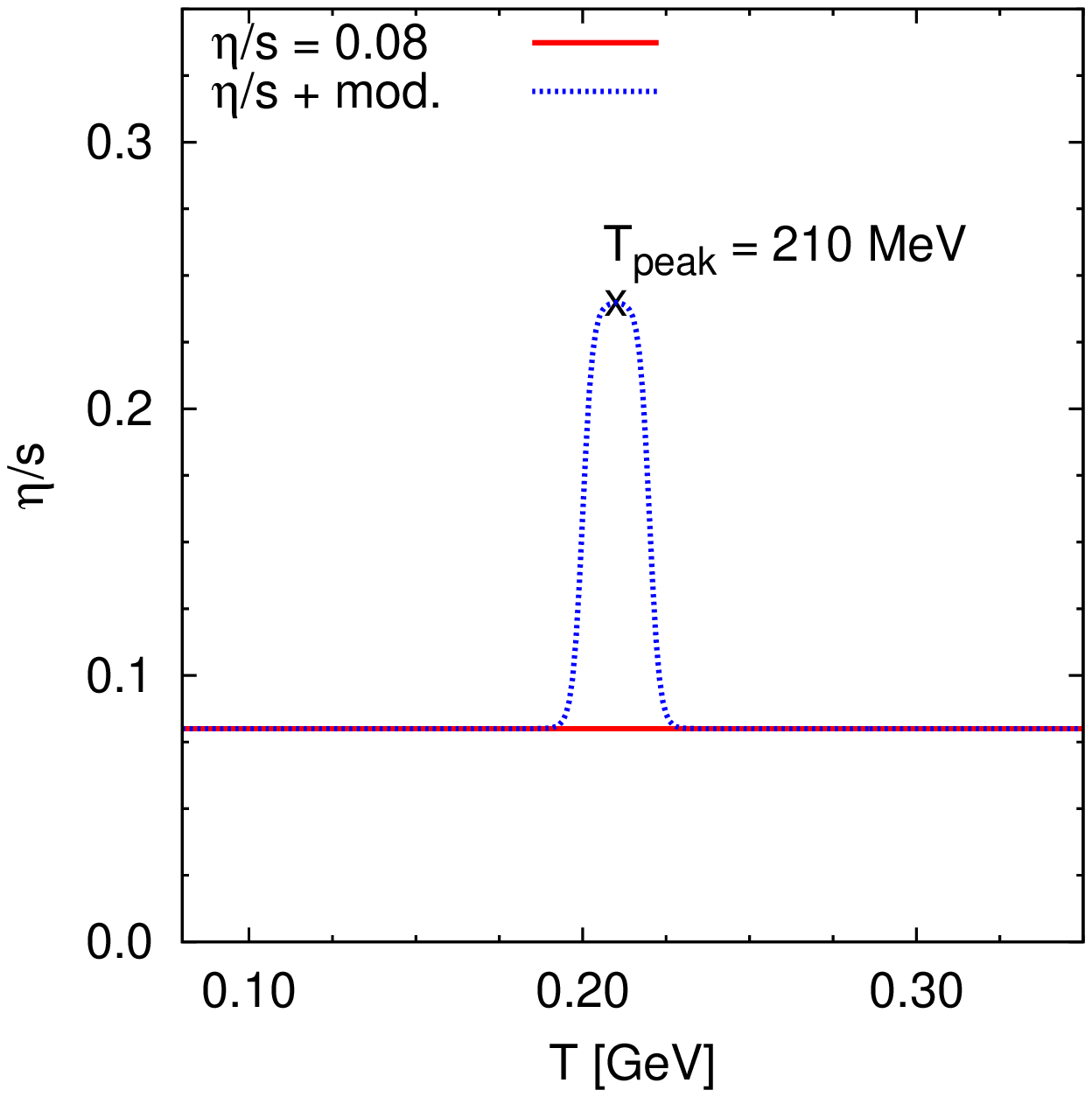}
  \caption{One of the $\eta/s$ parametrizations to find when
           flow anisotropies are most sensitive to $\eta/s$.}
  \label{fig:etaperspeak}
 \end{minipage}
  \hfill
 \begin{minipage}[t]{65mm}
  \includegraphics[width=65mm]{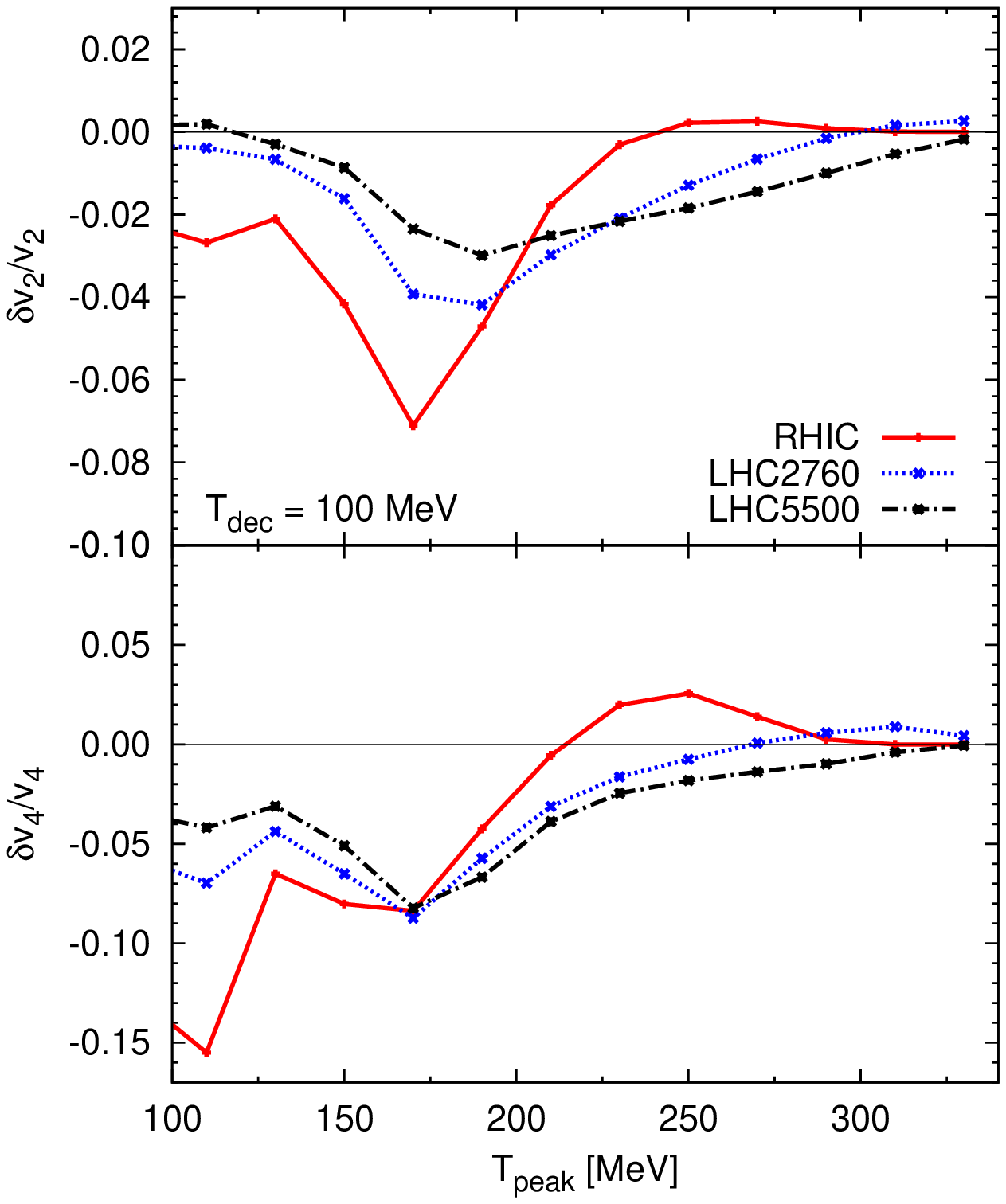}
 \caption{The change in the $p_T$-averaged $v_2$ and $v_4$ when there
   is a peak in the $\eta/s$ ratio at $T_\mathrm{peak}$.}
  \label{fig:peaks}
 \end{minipage}
\end{figure}

To further investigate the sensitivity of $v_2$ and $v_4$ to viscosity
at different temperatures, we devised a set of simple parametrizations
for $\eta/s(T)$, see Fig.~\ref{fig:etaperspeak}.  We take
$\eta/s=0.08$, except in the vicinity of a temperature
$T_\mathrm{peak}$, where $\eta/s(T_\mathrm{peak}) = 0.24$, and the
width of the peak is 10 MeV.  By varying $T_\mathrm{peak}$, we can
study at what temperature $\eta/s$ affects $v_2$ and $v_4$ most.

The results are shown in Fig.~\ref{fig:peaks} as the relative change
with respect to $v_2$ and $v_4$ evaluated using a constant
$\eta/s = 0.08$ during the entire evolution. It is seen that at RHIC,
the viscosity around $T=170$ MeV has the largest effect on $v_2$, but
the region of strongest sensitivity moves to larger temperatures and
becomes wider with increasing collision energy. If we ignore the
point at $T=110$ MeV (because its main effect is via $\delta f$), the
behaviour of $v_4$ is slightly different: With increasing collision
energy the suppression at large temperatures increases, and at low
temperatures decreases, but the temperature where $\eta/s$ suppresses
$v_4$ most hardly changes.

An interesting feature is that there is a region where $\eta/s$ does
not suppress flow anisotropies, but the larger the $\eta/s$ the larger
the anisotropy! This can be understood in the following way: At early
times the main effect of shear viscosity is to inhibit the
longitudinal expansion, and enhance the transverse expansion, instead
of reducing the difference between the expansion in in-plane and
out-of-plane directions. Thus, shear viscosity leads to larger
transverse flow velocity. A simple blast wave model~\cite{blast}
demonstrates that if nothing else changes, a larger transverse flow
velocity leads to a larger $v_2(p_T)$ of light particles. At RHIC a
similar reasoning leads to the insensitivity to the plasma viscosity,
since the effects of increasing flow velocity and smaller difference
between in-plane and out-of-plane directions cancel each other.

\begin{figure}[b]
\begin{minipage}[b]{60mm}
 \includegraphics[width=60mm]{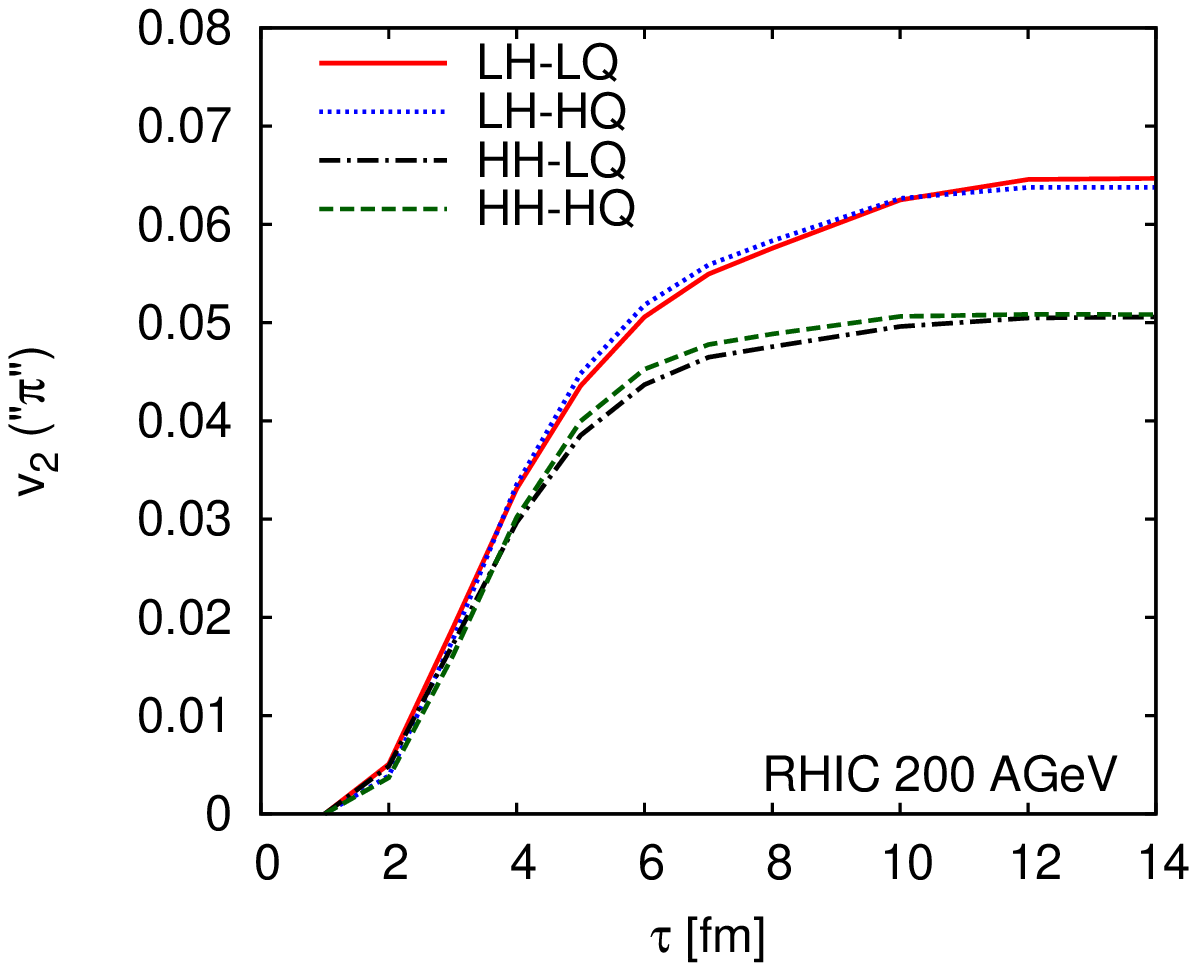}
\end{minipage}
 \hfill
\begin{minipage}[b]{60mm}
 \includegraphics[width=60mm]{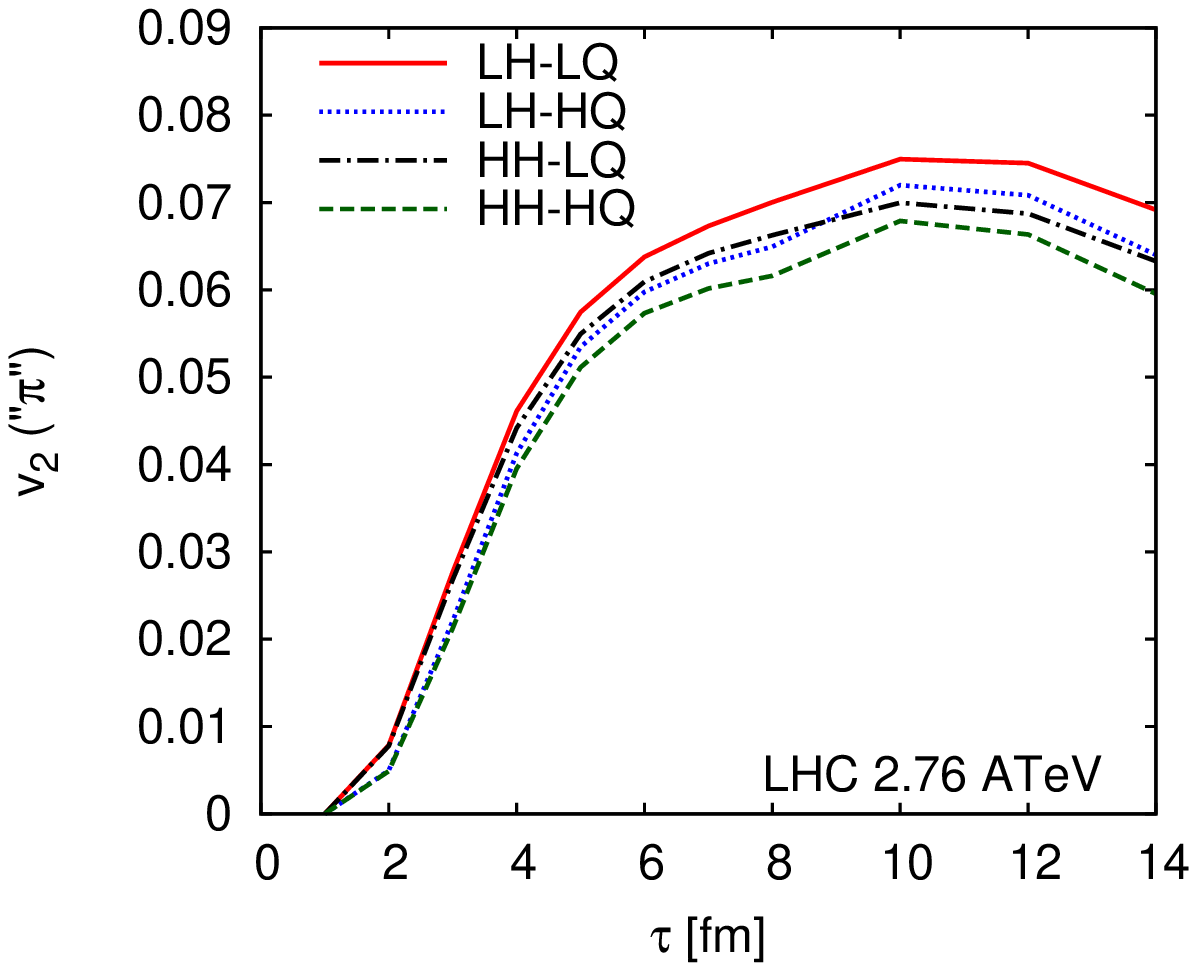}
\end{minipage}
\caption{Time-evolution of the $p_T$-averaged $v_2$ of fictitious
  $m=140$ MeV bosons (``$\pi$'') in 20-30\% most central Au+Au (left)
  and Pb+Pb (right) collisions at $\sqrt{s_{NN}} = 200$ GeV (RHIC) and
  $\sqrt{s_{NN}} = 2.76$ TeV (LHC), respectively.}
\label{fig:evolution}
\end{figure}

All this does not mean that $v_2$ would not be formed early. To take
into account the thermal motion during the evolution, we characterise
the time-evolution of $v_2$ by evaluating the $v_2$ of fictitious
$m=140$ MeV bosons at different times $\tau_i$. We use the
Cooper-Frye formula on a hypersurface consisting of two parts: A
constant temperature hypersurface with $T = T_\mathrm{dec}= 100$ MeV
and $\tau < \tau_i$, and a constant time hypersurface with
$\tau = \tau_i$ and $T > 100$ MeV. This approach has the advantage
that at the end of the evolution it matches the $v_2$ of thermal pions
without any adjustment.

As seen in Fig.~\ref{fig:evolution}, $v_2$ is built up early, but the
effect of $\eta/s$ at early times is relatively small. The left panel
depicts the evolution at RHIC, and one can see that at $\tau = 2$ fm,
the larger $\eta/s$ above the transition region has reduced the $v_2$
somewhat, but this difference is soon erased and the low-temperature
viscosity dominates the evolution from $\tau \sim 3$ fm on. The
situation is different in the evolution at the lower LHC energy
depicted in the right panel. Now the plasma viscosity causes a clear
difference at $\tau \sim 2$ fm, and this difference persists to much
later times than at RHIC. Eventually the hadronic viscosity takes over
and reorders the curves, but it cannot completely compensate the
differences built up earlier.

In summary, we have investigated how the temperature dependence of
$\eta/s$ affects the flow anisotropy coefficients $v_2$ and $v_4$. We
found that the temperature where $v_n$ is most sensitive to viscosity
varies with the collision energy---the larger the collision energy,
the larger the temperature where the suppression is strongest. We also
saw that $v_4$ is most sensitive to the viscosity at a lower
temperature than $v_2$. It remains to be seen whether this is a
general trend: the larger the $n$, the lower the temperature where
$v_n$ is most sensitive to the value of $\eta/s$.

\section*{Acknowledgement}

This work was supported by the Helmholtz International Center for FAIR
within the framework of the LOEWE program launched by the State of Hesse.
The work of 
H.N.\ was supported by the ExtreMe Matter Institute (EMMI).
P.H.\ is supported by BMBF under contract no.\ 06FY9092.
E.M.\ is supported by OTKA/NKTH 81655.

\end{document}